\documentstyle[aps,prc,epsfig]{revtex}

\begin{document}
\draft
\title{Measuring charge fluctuations \\
in high-energy nuclear collisions
}

\author{Stanis\l aw Mr\' owczy\' nski\footnote{Electronic address:
mrow@fuw.edu.pl}}

\address{So\l tan Institute for Nuclear Studies, \\
ul. Ho\.za 69, PL - 00-681 Warsaw, Poland \\
and Institute of Physics, \'Swi\c etokrzyska Academy, \\
ul. \'Swi\c etokrzyska 15, PL - 25-406 Kielce, Poland}

\date{1-st July 2002}

\maketitle

\begin{abstract}

Various measures of charge fluctuations in heavy-ion collisions
are discussed. Advantages of the $\Phi-$measure are 
demonstrated and its relation to other fluctuation measures   
is established. To get the relation, $\Phi$ is expressed through
the moments of multiplicity distribution. We study how the measures 
act in the case of a `background' model which represents the classical 
hadron gas in equilibrium. The model assumes statistical particle 
production constrained by charge conservation. It also takes 
into account both the effect of incomplete experimental apparatus 
acceptance and that of tracking inefficiency. The model is shown 
to approximately agree with the PHENIX and preliminary STAR data 
on the electric charge fluctuations. Finally, `background-free' 
measures are discussed.

\end{abstract}

\vspace{0.5cm}
 

\vspace{0.5cm}

\section{Introduction}

Fluctuations of strange, baryonic and electric charges studied on 
event-by-event basis have been repeatedly argued to provide dynamical 
information on high-energy heavy-ion collisions. Jeon and Koch 
\cite{Jeo99} have suggested to study the fluctuations of the 
ratio of positive to negative pions in order to measure the number 
of $\rho$ and $\omega$ resonances after hadronization. Gavin and 
Pruneau \cite{Gav99a} have found that the baryon number fluctuations 
are very sensitive to the degree of chemical equilibration of the 
systems produced in heavy-ion collisions at RHIC and LHC. Gavin with 
collaborators \cite{Gav99b,Bow01} have also suggested that the 
extraordinary baryon fluctuations can serve as a signal of the QCD 
tricritical point \cite{Hal98,Ste98,Ber99,Fod01}. Jeon and Koch 
\cite{Jeo00} and Asakwa, Heinz and M\"uller \cite{Asa00} have 
observed that the fluctuations of baryonic and electric charge are 
significantly smaller in the equilibrium quark-gluon plasma than in 
the hadron gas. Assuming that the fluctuations created in the quark 
phase survive the hadronization, the charge fluctuations normalized 
to the entropy, which is also assumed to be conserved, can be exploited 
as an indicator of the quark-gluon plasma formation in nuclear 
collisions \cite{Jeo00,Asa00}. The idea has been further discussed 
in \cite{Fia00,Fia01,Ble00,Hei00,Gaz00,Bop01,Shu01,Lin01}.

The NA49 measurement \cite{Rol98,Afa01} of the $K/\pi$ ratio 
at the SPS collision energy is somewhat discouraging. It suggests that 
the fluctuations in the central collisions are mostly of trivial 
statistical character. The conclusion has been theoretically 
analyzed in \cite{Jeo99,Bay99,Yan01}. The PHENIX \cite{Adc02} and 
preliminary STAR \cite{Vol01} results on charge fluctuations show 
that statistical noise also dominates at RHIC energies. Therefore,  
one faces a problem how to extract a small contribution of `dynamical' 
fluctuations of interest from the statistical background which, 
unfortunately, strongly depends on the collision centrality. 
Among other methods, the problem can be solved by means of the so-called 
$\Phi-$measure \cite{Gaz92} which has been successfully applied to 
the $p_T-$fluctuations \cite{App99}. $\Phi$ equals zero when 
inter-particle correlations are absent. It also eliminates 
`geometrical' fluctuations due to the impact parameter variation. 
Therefore, $\Phi$ is `deaf' to the statistical noise and `blind' 
to the collision centrality. 

Applicability of $\Phi$ to the fluctuations of chemical composition 
of the hadronic system produced in nuclear collisions has been 
already discussed in \cite{Gaz99,Mro99a}. In this paper we advocate
usefulness of $\Phi$ in studies of charge fluctuations which are 
obviously related to the chemical fluctuations. Advantages of $\Phi$
in such studies have been already demonstrated in a very recent paper 
\cite{Zar01}. Here, we also discuss another measure denoted as $\Gamma$, 
which is closely related to $\Phi$. However, $\Gamma$ is sensitive not 
only to the dynamical fluctuations, as is $\Phi$, but to the 
statistical fluctuations as well. We express $\Phi$ and $\Gamma$ 
through the moments of multiplicity distributions and then 
we compare them to the fluctuation measures suggested by other authors 
\cite{Jeo99,Jeo00,Asa00,Bay99,Vol01}. We also compute the measures
for a `background' model where the particle production is mostly 
statistical but constrained by charge conservation. The model 
represents the classical hadron gas in equilibrium, where hadron 
resonances are neglected. After taking into account a finite 
detector acceptance, which strongly reduces the effect of charge 
conservation, the model's predictions are compared to the PHENIX
\cite{Adc02} and preliminary STAR \cite{Vol01} data on the electric
fluctuations \cite{Vol01}. At the end, we discuss the measures where 
the background fluctuations, i.e., those given by the `background'
model, are eliminated. The measures are free of trivial effects 
caused by the charge conservation and finite detector acceptance. 

\section{$\Phi-$ and $\Gamma-$measure}

Let us introduce the measure $\Phi$ which describes the correlations 
(or fluctuations) of a single particle variable $x$. Here, $x$ is 
identified with the particle electric, baryonic or any other charge $q$. 
One defines a single-particle variable 
$z \buildrel \rm def \over = x - \overline{x}$ with the overbar 
denoting averaging over a single particle inclusive distribution. 
One easily observes that $\overline{z} = 0$. Further, we introduce 
the event variable $Z$, which is a multiparticle analog of $z$, 
defined as 
$Z \buildrel \rm def \over = \sum_{i=1}^{N}(x_i - \overline{x})$, 
where the summation runs over particles in a given event.
By construction, $\langle Z \rangle = 0$, where $\langle ... \rangle$ 
represents averaging over events. Finally, the $\Phi-$measure is defined 
in the following way:
\begin{eqnarray}\label{Phi}
\Phi \buildrel \rm def \over = 
\sqrt{\langle Z^2 \rangle \over \langle N \rangle } -
\sqrt{\overline{z^2}} \;.
\end{eqnarray}
It is evident that $\Phi = 0$, when no inter-particle correlations 
are present. The measure also possesses a less trivial property. 
Namely, $\Phi$ is {\it independent} of the distribution of the number 
of particle sources if the sources are identical and independent from each 
other \cite{Gaz92,Mro99b}. Thus, the $\Phi-$measure is `blind' to the 
impact parameter variation as long as the `physics' does not change 
with the collision centrality. In particular, the $\Phi$ is independent 
of the impact parameter if the nucleus-nucleus collision is a simple 
superposition of nucleon-nucleon interactions. 

As in the case of chemical fluctuations \cite{Mro99a}, we are going 
to express $\Phi$ through the moments of multiplicity distribution. 
Then, $\Phi$ can be compared to other fluctuation measures which
are usually defined in this way. We first consider a system of particles 
with two different values of charge. In principle, the system might be 
multi-component but only two charged components are taken into account. 
Then, $x$ equals either $q_1$ or $q_2$. The inclusive averages of $x$ 
and $x^2$ read
$$
\overline{x} = q_1P_1 + q_2 P_2\;, 
\;\;\;\;\;\;\;\;\;\;\;\;\;\;\;\;\;
\overline{x^2} = q_1^2P_1 +q_2^2 P_2 \;,
$$
where the probabilities to find a particle with $q_1$ and $q_2$, 
respectively, are
$$
P_i = { \langle N_i \rangle \over \langle N \rangle } 
\;,\;\;\;\;\;\;\; i=1,2
$$
with $N_i$ being the number of particles with charge $q_i$
and $N \equiv N_1 + N_2$. One easily finds that 
\begin{eqnarray}\label{222-2}
\overline{z^2} =  (q_1 - q_2)^2
{ \langle N_1 \rangle \langle N_2 \rangle \over 
\langle N \rangle^2 } \;.
\end{eqnarray}
Using the relation
$$
Z = Q - { \langle Q \rangle \over \langle N \rangle }\, N \;,
$$
where $Q = q_1N_1 + q_2N_2 $ is the system charge, we get 
$\langle Z \rangle =0$ and 
\begin{eqnarray}\label{111-2}
{\langle Z^2 \rangle \over \langle N \rangle} 
= {(q_1 -q_2)^2 \over \langle N \rangle^3}
\Big[ \langle N_1 \rangle^2 \langle N_2^2 \rangle
+\langle N_1^2 \rangle \langle N_2\rangle^2
- 2 \,\langle N_1 \rangle \langle N_2 \rangle 
\langle N_1 N_2 \rangle  \Big] \;,
\end{eqnarray}
which can be rewritten as
\begin{eqnarray}\label{111-2a}
{\langle Z^2 \rangle \over \langle N \rangle} = 
(q_1 -q_2)^2
{\langle N_1 \rangle^2 \langle N_2 \rangle^2 \over \langle N \rangle^3}  
\bigg[ 
{\langle N_1^2\rangle -\langle N_1\rangle^2 \over \langle N_1\rangle^2}
&+& 
{\langle N_2^2\rangle -\langle N_2\rangle^2 \over \langle N_2\rangle^2}
\\ [3mm] \nonumber
&-& 
2 {\langle N_1 N_2 \rangle - \langle N_1 \rangle \langle N_2 \rangle  
\over \langle N_1 \rangle \langle N_2 \rangle}
\bigg] \;.
\end{eqnarray}
The fluctuation measure $\Phi$ is completely determined by 
Eqs. (\ref{222-2}, \ref{111-2}). If the particle distributions are 
poissonian and independent from each other i.e. 
\begin{eqnarray}\label{poisson}
\langle N_i^2 \rangle - \langle N_i \rangle^2 
&=& \langle N_i \rangle \;,\;\;\;\;\;\;i=1,2
\\[3mm] \nonumber
\langle N_1 N_2 \rangle &=& \langle N_1 \rangle \langle N_2 \rangle \;,
\end{eqnarray}
one notices that 
\begin{eqnarray*}
{\langle Z^2 \rangle \over \langle N \rangle} 
= (q_1 - q_2)^2
{\langle N_1 \rangle \langle N_2 \rangle \over 
\langle N \rangle^2} 
\end{eqnarray*}
and $\Phi=0$. In general, $\Phi$ vanishes when particles are
independent form each other and the event's charge per particle 
is independent of the event's multiplicity. 

If one studies a system with particles carrying more than two 
different values of a given charge, Eqs.~(\ref{222-2},\ref{111-2}) 
have to be generalized. We first consider the generalization to 
the three-component system, such as that of positive ($q=1$), 
negative ($q=-1$) and neutral ($q=0$) hadrons. Although the neutral
particles do not contribute to the system's charge, they do contribute
to the fluctuations measured by $\Phi$. After rather lengthy 
calculations one finds
\begin{eqnarray}\label{222-3}
\overline{z^2} = (q_1 - q_2)^2
{ \langle N_1 \rangle \langle N_2 \rangle \over 
\langle N \rangle^2 }
+ (q_1 - q_3)^2
{ \langle N_1 \rangle \langle N_3 \rangle \over 
\langle N \rangle^2 }
+(q_2 - q_3)^2
{ \langle N_2 \rangle \langle N_3 \rangle \over 
\langle N \rangle^2 }
\end{eqnarray}
and 
\begin{eqnarray}\label{111-3}
{\langle Z^2 \rangle \over \langle N \rangle} 
= {(q_1 -q_2)^2 \over \langle N \rangle^3} A_{12}
+ {(q_1 -q_3)^2 \over \langle N \rangle^3} A_{13}
+ {(q_2 -q_3)^2 \over \langle N \rangle^3} A_{23} \;,
\end{eqnarray}
where $N \equiv N_1 + N_2 + N_3$ and 
\begin{eqnarray}\label{A} 
A_{12} &\equiv&
\langle N_1^2 \rangle \big(
\langle N_2 \rangle^2 + \langle N_2 \rangle \langle N_3 \rangle \big)
+ \langle N_2^2 \rangle \big(
\langle N_1 \rangle^2 + \langle N_1 \rangle \langle N_3 \rangle \big)
-\langle N_1 \rangle \langle N_2 \rangle \langle N_3^2 \rangle \\[2mm]
\nonumber
&-&\langle N_1 N_2 \rangle \big(
2\,\langle N_1 \rangle \langle N_2 \rangle +\langle N_2 \rangle \langle N_3 \rangle
+\langle N_1 \rangle \langle N_3 \rangle +\langle N_3 \rangle^2 \rangle \big) \\[2mm]
\nonumber
&+&\langle N_2 N_3 \rangle \big(
\langle N_1 \rangle^2  - \langle N_1 \rangle \langle N_2 \rangle
+\langle N_1 \rangle \langle N_3 \rangle  \big) 
+\langle N_1 N_3 \rangle \big(
\langle N_2 \rangle^2  - \langle N_1 \rangle \langle N_2 \rangle
+\langle N_2 \rangle \langle N_3 \rangle  \big) \;.
\end{eqnarray}
$A_{13}$ can be found from $A_{12}$ by swapping indices 
$2 \leftrightarrow 3$ and $A_{23}$ coincides with $A_{13}$ when 
$1 \leftrightarrow 2$. One easily shows that for the poissonian distribution 
(\ref{poisson}), $\langle Z^2 \rangle / \langle N \rangle = \overline{z^2}$ 
and $\Phi=0$. 

The formulas (\ref{222-3},\ref{111-3},\ref{A}) can be further 
generalized to a system of four or higher number of particle species, 
such as the quark-gluon plasma. While the modifications of 
Eqs.~(\ref{222-3},\ref{111-3}) are obvious, Eq.~(\ref{A}) should be 
understood in such a way that $N_3$ represents all particles other 
than those carrying charges $q_1$ or $q_2$. 

The $\Phi-$measure has been designed to look for `dynamical' 
fluctuations. As seen, it vanishes when the fluctuations are of 
simple statistical origin. However, for the theoretical suggestion
\cite{Jeo00,Asa00} the fluctuation {\it magnitude} is the main issue. 
It has been observed by the authors of \cite{Jeo00,Asa00} that the 
statistical fluctuations generated in the quark-gluon phase 
are significantly smaller than those in the hadron gas. Therefore,
if the quark-gluon plasma fluctuations are frozen due to the fast
longitudinal expansion of the system the fluctuations observed 
at the hadron phase are significantly smaller than the statistical
fluctuations characteristic for the hadron gas. Then, instead
of the $\Phi-$measure defined by Eq.~(\ref{Phi}), one cane use 
$\langle Z^2 \rangle / \langle N \rangle$, which is, as is $\Phi$,
insensitive to the distribution of the independent particle sources. 
Therefore, we define a measure
\begin{equation}\label{G}
\Gamma \buildrel \rm def \over = {1 \over \langle N \rangle}
\Big\langle \Big(Q - {\langle Q \rangle \over \langle N \rangle} 
\; N \Big)^2 \Big\rangle 
=  {\langle Z^2 \rangle \over \langle N \rangle} \;,
\end{equation} 
which was introduced in the very first paper on $\Phi$ 
\cite{Gaz92}. As discussed below, $\Gamma$, which measures both 
the dynamical and statistical fluctuations, can be very useful 
in the experimental data analysis.

\section{Other measures}

In this section we compare $\Phi$ and $\Gamma$ to other measures of charge 
fluctuations. We limit the comparison to the case of two-component
system. The charge fluctuations can be studied by means of the ratio
of the multiplicities of particles of different charges, $R = N_1/N_2$
\cite{Jeo99,Jeo00}. One finds \cite{Jeo99,Bay99} 
that to the second order in the fluctuations of numbers of particles
\begin{equation}\label{R-fluc}
\langle R^2 \rangle - \langle R \rangle^2 \cong 
{\langle N_1 \rangle^2 \over \langle N_2 \rangle^2}
\Bigg[
{\langle N_1^2 \rangle - \langle N_1 \rangle^2 \over \langle N_1 \rangle^2}
+ {\langle N_2^2 \rangle - \langle N_2 \rangle^2 \over \langle N_2 \rangle^2}
-2\, {\langle N_1N_2 \rangle - \langle N_1 \rangle \langle N_2 \rangle 
\over \langle N_1 \rangle \langle N_2 \rangle } \Bigg] \;.
\end{equation}
Instead of $R$, one can use $F = Q/N$ where $Q$ is, as before,
the system charge \cite{Jeo00}. If we deal with the particles of opposite
unit charges $(q_1 = - q_2 = 1)$, $R$ and $F$ are simply related to
each other. Specifically,
$$
R= {1 + F \over  1 - F} = 1 + 2F + 2F^2 + {\cal O}(F^3) \;.
$$
Consequently,
$$
\langle R^2 \rangle - \langle R \rangle^2 \cong
4 \big(\langle F^2 \rangle - \langle F \rangle^2 \big) \;.
$$

Comparing Eqs.~(\ref{111-2a},\ref{R-fluc}) to each other, we get the
relation 
\begin{equation}\label{G-D}
\Gamma = 
{\langle Z^2 \rangle \over \langle N \rangle} 
\cong (q_1 - q_2)^2 {\langle N_2 \rangle^4 \over \langle N \rangle^3 }
\Big[ \langle R^2 \rangle - \langle R \rangle^2 \Big] 
= (q_1 - q_2)^2 {\langle N_2 \rangle^4 \over \langle N \rangle^4 } \; D \;,
\end{equation}
where
\begin{equation}\label{D}
D \buildrel \rm def \over = \langle N \rangle 
\big[ \langle R^2 \rangle - \langle R \rangle^2 \big]
\end{equation}
is the charge fluctuation measure advocated in \cite{Jeo00,Ble00}.
Since Eq.~(\ref{R-fluc}) holds only for sufficiently small fluctuations
the same is true for the relationship (\ref{G-D}). Therefore, $D$ is
independent of the particle source distribution for small fluctuations
only while $\Gamma$ possesses this property for fluctuations of any 
size.

Another natural measure of charge fluctuations is \cite{Jeo00,Asa00,Adc02}
$$
v(Q) \buildrel \rm def \over = 
{\langle Q^2 \rangle - \langle Q \rangle ^2 \over \langle N \rangle }=
{1\over \langle N \rangle}\,\Big[
  q_1^2 \big(\langle N_1^2 \rangle - \langle N_1 \rangle^2 \big)
+ q_2^2 \big(\langle N_2^2 \rangle - \langle N_2 \rangle^2 \big)
+ 2q_1q_2 \big(\langle N_1 N_2 \rangle 
- \langle N_1 \rangle \langle N_2 \rangle \big) \Big] \;,
$$
which is not simply related to $\Gamma$, except in the two special cases:
when $\langle Q \rangle = 0$ and when $N$ is fixed. Then, $\Gamma = v(Q)$.  
One also observes that for the poissonian distribution of $N_1$ and $N_2$
there is a relation:
\begin{equation}\label{Gamma-v} 
\Gamma = v(Q) - {\langle Q \rangle^2 \over \langle N \rangle^2} \;,  
\end{equation}    
which will be used in Sec. V to discuss the PHENIX data \cite{Adc02}.

One more measure has been proposed in the experimental study 
\cite{Vol01}. Namely,
\begin{equation}\label{nu}
\nu \buildrel \rm def \over = 
\Big\langle \Big({N_1 \over \langle N_1 \rangle} 
- {N_2 \over \langle N_2 \rangle} \Big)^2 \Big\rangle = 
  {\langle N_1^2 \rangle - \langle N_1 \rangle^2 \over \langle N_1 \rangle^2}
+ {\langle N_2^2 \rangle - \langle N_2 \rangle^2 \over \langle N_2 \rangle^2}
-2\, {\langle N_1N_2 \rangle - \langle N_1 \rangle \langle N_2 \rangle 
\over \langle N_1 \rangle \langle N_2 \rangle } \;.
\end{equation} 
The measure $\nu$, which is computed in the poissonian approximation (\ref{poisson}), has been called the `statistical' contribution and 
denoted as $\nu_{\rm stat}$;
\begin{equation}\label{nu-stat}
\nu_{\rm stat} = { 1 \over \langle N_1 \rangle} 
+ {1 \over \langle N_2 \rangle} 
= {\langle N \rangle \over \langle N_1 \rangle \langle N_2 \rangle} \;.
\end{equation} 
The authors of \cite{Vol01} has also used the ratio 
$\nu / \nu_{\rm stat}$ and the difference
\begin{equation}\label{nu-dyn}
\nu_{\rm dyn} = \nu - \nu_{\rm stat} \;,
\end{equation}
which they call the `dynamical' contribution.

Comparing Eqs.~(\ref{222-2},\ref{111-2a},\ref{G}) to 
Eqs.~(\ref{nu},\ref{nu-stat},\ref{nu-dyn}), one finds that
for $q_1=-q_2 = 1$
\begin{equation}\label{G-nu}
\Gamma = 4 
{\langle N_1 \rangle^2 \langle N_2 \rangle^2 \over \langle N \rangle^3}
\; \nu \;,
\end{equation}
\begin{equation}\label{z-nu}
\overline{z^2}  = 4 
{\langle N_1 \rangle^2 \langle N_2 \rangle^2 \over \langle N \rangle^3}
\; \nu_{\rm stat} \;,
\end{equation}
\begin{equation}\label{Phi-nu}
\Phi  = 2
{\langle N_1 \rangle \langle N_2 \rangle \over \langle N \rangle}
\bigg( \sqrt{{\nu \over \langle N \rangle}} 
    - \sqrt{{\nu_{\rm stat} \over \langle N \rangle}} \; \bigg)
\cong 
{\langle N_1 \rangle^{3/2}\langle N_2 \rangle^{3/2} 
\over \langle N \rangle^2 } \; \nu_{\rm dyn} \;.
\end{equation}
The second approximate equality in Eq.~(\ref{Phi-nu}) holds
for $\nu_{\rm stat} \gg \nu_{\rm dyn}$.

At the end of this section we remark that 
\begin{eqnarray*}
\Gamma = v(Q) = {1 \over 4} \; D = {\langle N \rangle \over 4} \; \nu = 1 \;,
\end{eqnarray*}
when $\langle Q \rangle = 0$, $q_1 = - q_2 = 1$ and the poissonian 
approximation (\ref{poisson}) holds. $\Phi$ and $\nu_{\rm dyn}$
vanish by definition in this case.

\section{`Background' model}

In this section we discuss a very simple model of charge fluctuations 
which takes into account the charge conservation and the fact that 
one usually observes only a fraction of all charged particles produced 
in nuclear collisions. We consider the charges of two values, say $+1$ 
and $-1$. The multiplicities of positive (negative) particles are denoted 
here as $N_+$ ($N_-$). At the beginning we assume that {\em all} charged particles are observed. Then, $N_+ - N_- = Q$, where $Q$ denotes the 
electric, baryonic or any other conserved charge which is the same for 
all events under consideration. The multiplicity distribution of
negative and positive particles can be written as
\begin{equation}\label{multi-distri1}
P_{N_+ N_-} = P_{N_-} \delta_{N_+}^{N_- + Q} \;,
\end{equation}
or 
\begin{equation}\label{multi-distri2}
P_{N_+N_-} = P_{N_+} \delta_{N_-}^{N_+ - Q} \;.
\end{equation}
Using the distribution (\ref{multi-distri1}) one immediately finds
\begin{eqnarray*}
\langle N_+ \rangle &=& \langle N_- \rangle + Q \;, \\
\langle N_+^2 \rangle - \langle N_+ \rangle^2 &=& 
\langle N_-^2 \rangle - \langle N_- \rangle^2 \;, \\
\langle N_+N_- \rangle - \langle N_+ \rangle \langle N_- \rangle
&=& \langle N_-^2 \rangle - \langle N_- \rangle^2 \;. 
\end{eqnarray*}
Then, Eqs.~(\ref{222-2},\ref{111-2}) give
\begin{eqnarray*}
\overline{z^2} &=&  4 
{ \langle N_- \rangle (\langle N_- \rangle +Q) \over 
\langle N \rangle^2 } \;, \\
\Gamma =  
{\langle Z^2 \rangle \over \langle N \rangle} 
&=& {4 Q^2 \over \langle N \rangle^3}
\big( \langle N_-^2 \rangle - \langle N_- \rangle^2 \big) \;,
\end{eqnarray*}
where $N \equiv N_+ + N_- = 2N_- + Q$. For $Q=0$ one gets
$\overline{z^2} = 1$, $\langle Z^2 \rangle / \langle N \rangle = 0$
and $\Phi = -1$.

Now, we consider a situation when only a fraction of charged particles
produced in nuclear collision is observed. As many other authors,
see e.g. \cite{Mah01}, we assume that every particle is registered
with the probability $p$ which is the same for all particles
independently of their momenta. The particle is lost with the
probability $(1-p)$. In this way, we model both the effect of tracking
inefficiency and inefficiency and inefficiency and incomplete detector
acceptance. Since the number of observed particles is given by the
binomial distribution the multiplicity distribution of observed
particles reads
\begin{equation}\label{multi-distri3}
{\cal P}_{N_+N_-} = \sum_{M_+=N_+}^{\infty} \sum_{M_-=N_-}^{\infty}
P_{M_+M_-} {M_+ \choose N_+} p^{N_+} (1 - p)^{M_+ - N_+} \;
{M_- \choose N_-} p^{N_-} (1 - p)^{M_- - N_-} \;,
\end{equation}
where $M_{\pm}$ corresponds to the produced while $N_{\pm}$ to the
observed particles.

\vspace{-0.7cm}
\begin{figure}
\centerline{\epsfig{file=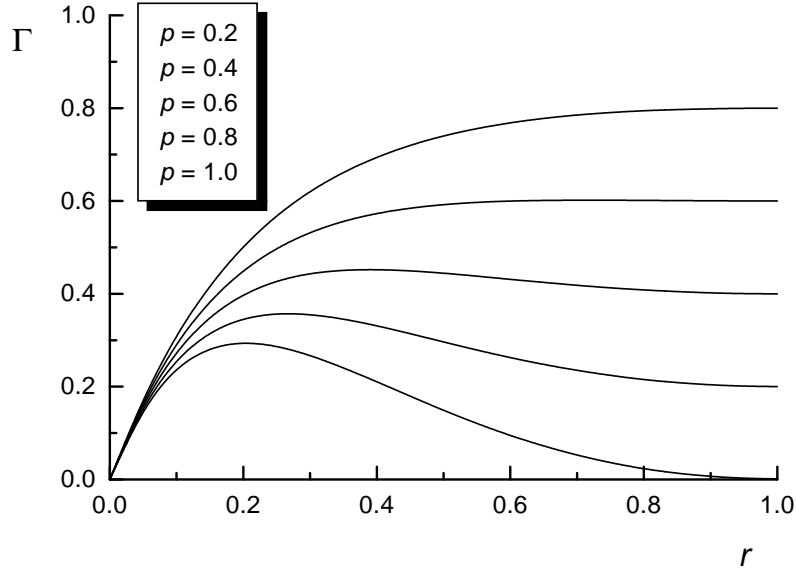,bbllx=14pt,bblly=440pt,bburx=540pt,bbury=850
pt,width=0.6\linewidth}}
\vspace{0.9cm}
\caption{$\Gamma$ as a function of the multiplicity ratio
$r\equiv \langle N_- \rangle / \langle N_+ \rangle$  for several
values of $p$. The top line corresponds to $p=0.2$,
the next one to $p=0.4$, etc.}
\end{figure}

Substituting the distribution (\ref{multi-distri1}) into 
(\ref{multi-distri3}) one gets
\begin{eqnarray*}
\langle N_- \rangle &=& p \langle M_- \rangle  \;, \\
\langle N_+ \rangle &=& p \langle M_- \rangle + pQ \;, \\
\langle N_-^2 \rangle - \langle N_- \rangle^2 &=&
p^2 \big( \langle M_-^2 \rangle - \langle M_- \rangle^2 \big)
+ (p-p^2)\langle M_- \rangle  \;, \\
\langle N_+^2 \rangle - \langle N_+ \rangle^2 &=& 
p^2 \big( \langle M_-^2 \rangle - \langle M_- \rangle^2 \big)
+ (p-p^2)\big( \langle M_- \rangle + Q \big)  \;, \\
\langle N_+N_- \rangle - \langle N_+ \rangle \langle N_- \rangle
&=&  p^2 \big( \langle M_-^2 \rangle - \langle M_- \rangle^2 \big) \;. 
\end{eqnarray*}

Our further considerations are limited to the poissonian approximation.
Namely, one assumes that $P_{N_-}$ in (\ref{multi-distri1}) is a Poisson
distribution as in a classical hadron gas in equilibrium. Then, one gets
\begin{eqnarray*}
\langle N_-^2 \rangle - \langle N_- \rangle^2 &=&
\langle N_- \rangle \;, \\
\langle N_+^2 \rangle - \langle N_+ \rangle^2 &=& 
\langle N_+ \rangle - p^2Q \;, \\
\langle N_+N_- \rangle - \langle N_+ \rangle \langle N_- \rangle
&=& p \langle N_- \rangle \;. 
\end{eqnarray*}
One can see here that the multiplicity distribution of negative particles 
is poissonian but that of positive ones is not. Substituting the above 
formulas to Eqs.~(\ref{222-2},\ref{111-2}) one gets
\begin{eqnarray}\label{final1a}
\overline{z^2} &=&  {4 r \over (1 + r)^2} \;, \\
\label{final2a}
\Gamma
&=& {4 r \over (1 + r)^2} - 4p \; {r^2 (3 - r) \over (1 + r)^3} \;.
\end{eqnarray}
where $r \equiv \langle N_- \rangle / \langle N_+ \rangle$.
When the system is symmetric i.e. $Q=0$ and consequently $r=1$, 
the formulas (\ref{final1a}, \ref{final2a}) simplify to 
$$
\overline{z^2} =  1 \;, \;\;\;\;\;\;\;\;\;\;
\Gamma
= 1- p \;,
$$
and give
\begin{equation}\label{Phi-back}
\Phi = \sqrt{1-p} - 1\;.
\end{equation}
We note here that Eq.~(\ref{Phi-back}) holds not only for
the poissonian approximation but for any distribution 
(\ref{multi-distri1}) with $Q=0$.

\vspace{-0.8cm}
\begin{figure}
\centerline{\epsfig{file=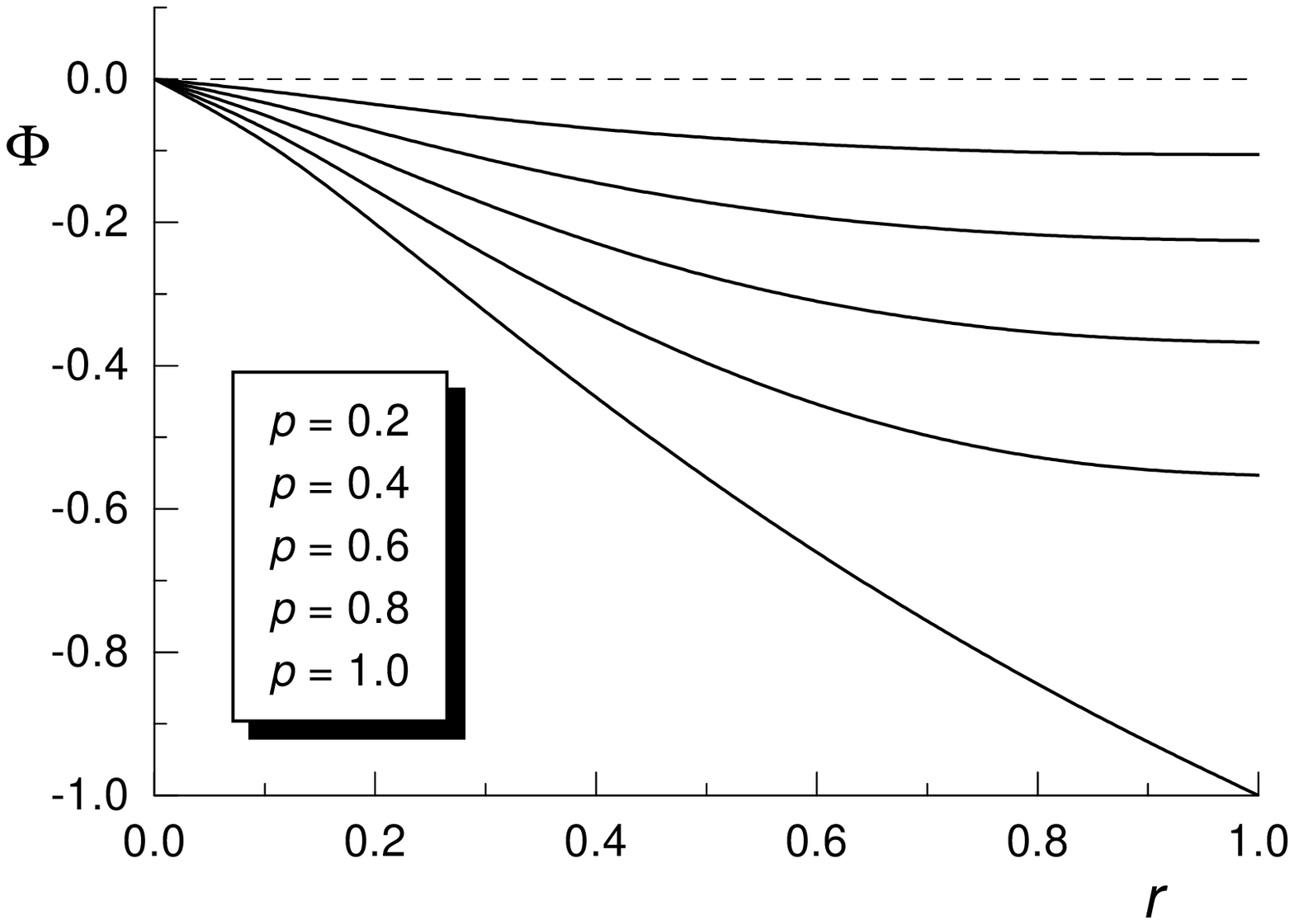,bbllx=14pt,bblly=440pt,bburx=540pt,bbury=850
pt,width=0.6\linewidth}}
\vspace{0.3cm}
\caption{$\Phi $ as a function of the multiplicity ratio
$r\equiv \langle N_- \rangle / \langle N_+ \rangle$  for several
values of $p$. The top (solid) line corresponds to $p=0.2$,
the next one to $p=0.4$, etc.}
\end{figure}

\vspace{-0.5cm}
\begin{figure}
\centerline{\epsfig{file=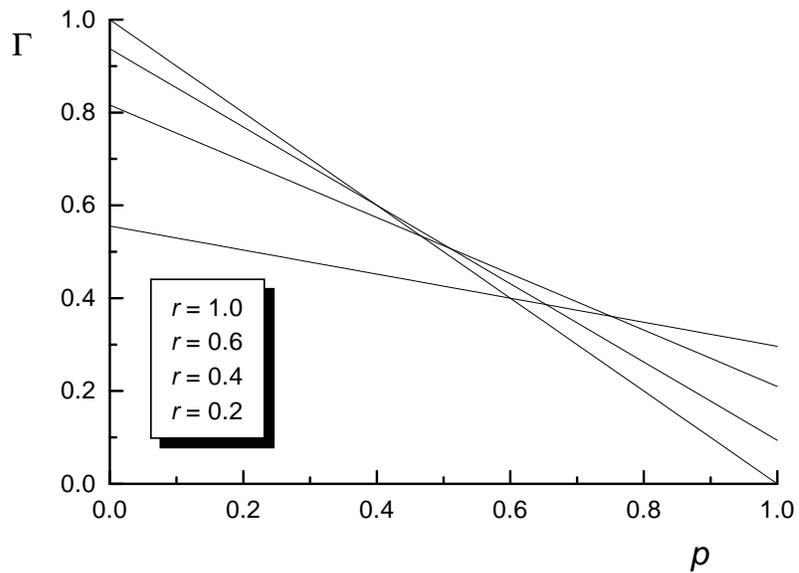,bbllx=14pt,bblly=440pt,bburx=540pt,bbury=850
pt,width=0.6\linewidth}}
\vspace{0.9cm}
\caption{$\Gamma$ as a function of the particle registration
probability $p$ for several values of
$r\equiv \langle N_- \rangle / \langle N_+ \rangle$.
The steepest line corresponds to $r = 1.0$, the second steepest
to $r=0.6$, etc. }
\end{figure}

The poissonian approximation can also be implemented in such
a way that $P_{N_+}$ in (\ref{multi-distri2}) is a Poisson
distribution. Then, the multiplicity distribution of positive
particles is poissonian while that of the negative ones is not.
In this case the final results are
\begin{eqnarray}\label{final1b}
\overline{z^2} &=&  {4 r \over (1 + r)^2} \;, \\
\label{final2b}
\Gamma
&=& {4 r \over (1 + r)^2} - 4p \; {3 r - 1 \over (1 + r)^3} \;.
\end{eqnarray}
Obviously, the results (\ref{final2a}) and (\ref{final2b}) differ
from each other. However, one observes that Eq.~(\ref{final2a}) holds
for $Q \ge 0$ ($r \le 1$) while Eq.~(\ref{final2b}) for $Q \le 0$
($r \ge 1$). Otherwise the multiplicity distributions which are
assumed to be poissonian cannot be poissonian because they must
vanish for $N_{\pm} < |Q|$. One further observes that
Eqs.~(\ref{final1a}, \ref{final2a}) changes into
Eqs.~(\ref{final1b}, \ref{final2b}) under the transformation
$r \rightarrow 1/r$. We also note that our final results i.e.
Eqs.~(\ref{final1a}, \ref{final2a}) or
Eqs.~(\ref{final1b}, \ref{final2b}) depend on $Q$ only
through $r$. Therefore, the initial assumption that $Q$ is
the same for all events can be relaxed and the events of
different $Q$ but of the same $r$ can be combined.

\vspace{-0.5cm}
\begin{figure}
\centerline{\epsfig{file=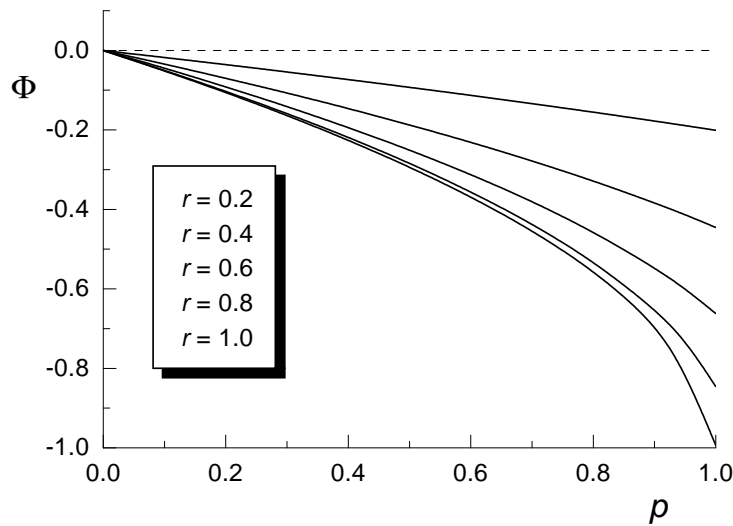,bbllx=14pt,bblly=440pt,bburx=540pt,bbury=850
pt,width=0.6\linewidth}}
\vspace{0.3cm}
\caption{$\Phi$ as a function of the particle registration
probability $p$ for several values of
$r\equiv \langle N_- \rangle / \langle N_+ \rangle$.
The top (solid) line corresponds to $r=0.2$, the next one
to $r=0.4$, etc. }
\end{figure}

The $\Gamma-$ and $\Phi-$measure given by 
Eqs.(\ref{final1a}, \ref{final2a}) are shown in Figs. 1-4.
As seen, the fluctuations measured by $\Gamma$ are suppressed
by the charge conservation when $p \rightarrow 1$. However,
the suppression is complete for $r=1$ only. When the net charge is
nonzero ($r < 1$) the fluctuations occur even at $p=1$. The same
is true for the measures $D$ and $\nu$. Within the `background' model 
$\Phi$ is determined by charge conservation. Therefore, the absolute 
value of $\Phi$ is the largest when $p=1$ and $r=1$. Then, the system
is constrained most effectively. When $r \rightarrow 0$ and
$p \rightarrow 0$ the effect of charge conservation is diluted.

Experimental data are often contaminated by particles coming from
secondary interactions in the detector material or, in general, from 
sources which are different than the interactions under study. 
These background particles, which are not influenced by charge  
conservation discussed above, should be also included in our
`background' model.  We assume that multiplicity distributions
of both positive and negative background particles are poissonian
and that their average multiplicities equal to each other. Then,
one finds that Eq.~(\ref{final1a}) remains unchanged while  
 Eq.~(\ref{final2a}) is modified as
\begin{eqnarray}\label{final2c}
\Gamma
= {4 r \over (1 + r)^2} - 4p \; {r^2 (3 - r - 2b) \over (1 + r)^3} \;.
\end{eqnarray}
where $b$ is the fraction of background particles among all negative
ones. When $r=1$ Eq.~(\ref{final2c}) simplifies to $\Gamma = 1 - p(1-b)$.

Using Eqs.~(\ref{G-nu},\ref{z-nu}) and (\ref{final1a},\ref{final2a})
one finds that the ratio $\nu /\nu_{\rm stat}$ studied by
the STAR collaboration \cite{Vol01} equals
\begin{equation} \label{nu/nu}
{\nu \over \nu_{\rm stat}} = {\Gamma \over \overline{z^2}}
= 1 - p \; {r (3 -r) \over 1 + r} \;.
\end{equation}
In the next section Eqs~(\ref{final2c},\ref{nu/nu}) are confronted with the
experimental data.

\section{Experimental data}

As noted in the Introduction, there are PHENIX \cite{Adc02} and 
STAR \cite{Vol01} measurements of the electric charge fluctuations  
in Au-Au collisions at $\sqrt{s_{NN}} = 130$ GeV. The result reported 
by PHENIX collaboration is 
$v(Q) = 0.965 \pm 0.007 \;({\rm stat.}) - 0.019 \;({\rm syst.})$.
It corresponds to 10\% most central collisions but the data
show no significant centrality dependence. The measurements have
been performed in the pseudorapidity, transverse momentum and
azimuthal angle regions: $-0.35 < \eta < 0.35$, $0.2 \;{\rm GeV}/c < p_T$
and $0 < \phi < \pi/2$. The amount of background particles, which
come from sources different than the Au-Au interactions of interest,
has been estimated as 20\% ($b=0.2$). 
 
According to the  preliminary STAR results the ratio 
$\nu /\nu_{\rm stat}$ is within experimental errors independent 
of centrality and equals $0.80 \pm 0.03$ for $-0.5 < \eta < 0.5$,
$0.1 < p_T < 2.0 \; {\rm GeV}/c$ and full azimuthal angle coverage.
The background is of order of 1\% in the case of STAR and it is
further neglected in our considerations. To compare the experimental 
results to the `background' model predictions one needs to estimate 
two parameters: $p$ and $r$.

The BRAHMS collaboration has found \cite{Bea01} that on average 
$3860 \pm 300$ charged particles (within $-4.7 < \eta < 4.7$) are 
produced in the most central Au-Au collisions with 352 participants. 
Among these particles $553 \pm 36 \;$ i.e. $14 \pm 2$ \% appears 
in the interval $-0.5 < \eta < 0.5$. The corresponding number 
is $10 \pm 1.4$ \% for $-0.35 < \eta < 0.35$.
Using the exponential parameterization of 
the transverse momentum distribution ($\sim p_T e^{-p_T/T}$), 
we have also estimated that about 10\% and 26\% of particles
for STAR and PHENIX, respectively, are lost because of the low 
$p_T$ cut-off $p^{\rm min}_T = 100 \; {\rm MeV} \cong T/2$
and $p^{\rm min}_T = 200 \; {\rm MeV} \cong T$.  Finally, taking
into account the tracking efficiency, which is about 90\% for STAR
and $80 \pm 5$\% for PHENIX, we have obtained $p = 0.11 \pm 0.02$
(STAR) and $p=0.015 \pm 0.003$ (PHENIX). In the case of PHENIX,
an additional factor 0.25 has been included due to the limited
azimuthal coverage.

As already mentioned, according to the BRAHMS data \cite{Bea01}, 
there is $3860 \pm 300$ charged particles produced by 352 
participants. This number corresponds, on average, to 140 
protons and 212 neutrons in Au-Au collisions. Therefore, 
$\langle N_+ \rangle = 2000 \pm 150$ and
$\langle N_- \rangle = 1860 \pm 150$. Consequently, 
$r = \langle N_- \rangle / \langle N_+ \rangle = 0.93 \pm 0.14$.

Substituting the estimated values of $p$ and $r$ into 
Eqs.~(\ref{final2c},\ref{nu/nu}) and using Eq.~(\ref{Gamma-v}) 
one finds $v(Q) = 0.990 \pm 0.003$ for PHENIX and 
and $\nu /\nu_{\rm stat} = 0.89 \pm 0.02$ for STAR. Both
theoreical estimates are somewhat higher than the experimental
values: $0.965+0.007-0.026$ (PHENIX) and $0.80 \pm 0.03$ (STAR).
The small differences are presumably due to the neutral
resonances which decay into charge hadrons and effectively reduce 
the charge fluctuations \cite{Jeo00,Asa00}. As shown by 
Zaranek \cite{Zar01}, the effect of resonances strongly depends 
on the rapidity window, where the charge hadrons are observed, 
and the fluctuations can be even enhanced in sufficiently small 
windows. In any case, we conclude this section by saying that the 
model of a classical hadron gas in equilibrium approximately explains 
the experimentaly observed electric charge fluctuations and there
is not much space for dynamical effects.

\section{Background free measures}

Since the effects of charge conservation and incomplete acceptance 
are of no real interest, it is desirable to use such measures which
are insensitive to both effects. The authors of Ref. \cite{Ble00}
introduced the modified $D-$measure which in our notation equals 
$$
\tilde D = { 1 \over r^2 (1 - p)} \; D \;,
$$
where $D$ is defined by Eq.~(\ref{D}). Using the relation (\ref{G-D}) 
and Eq.~(\ref{final2a}) one finds $\tilde D$ corresponding to the 
`background' model. It equals
\begin{equation}\label{D-tilde}
\tilde D = {(1 + r)^2 \over r (1 - p)} \; 
\Big[ 1 - p \; { r (3-r) \over 1 +r} \Big] \;.
\end{equation}
As seen in Eq.~(\ref{D-tilde}), $\tilde D \not= 4$ for 
$r \not= 1$. Figs. 5 and 6, where Eq.~(\ref{D-tilde}) is illustrated, 
show that the difference can be significant. A similar conclusion
has been recently drawn by Zaranek \cite{Zar01} who has studied how 
$\tilde D$ behaves in variety of simple fluctuation models. He has
also proposed another background-free measure 
$\Delta \Phi = \Phi - \Phi_0$ with $\Phi_0$ given by 
Eq.~(\ref{Phi-back}) which holds for $Q=0$. Since one often deals
with the systems where $Q > 0$ it would be preferable to use as $\Phi_0$
the expression given by Eqs. (\ref{final1a},\ref{final2a}).

When the electric charge fluctuations are studied, $r$ is close
to unity at sufficiently high collision energies because the
multiplicity of the produced charged hadrons is significantly
larger than the number of participating protons. Then, the
$(1 - p)$ correction \cite{Ble00} works well. At lower energies,
$r$ significantly differs from 1. The ratio is also noticeably
smaller than unity when the baryon number fluctuations are
studied. Then, one can use our `background' model to construct
a variety of the background free measures. For example,
$\Gamma - \Gamma_0$ or $\Gamma/\Gamma_0$, where $\Gamma_0$ is
given by Eq.~(\ref{final2a}).

\vspace{-0.7cm}
\begin{figure}
\centerline{\epsfig{file=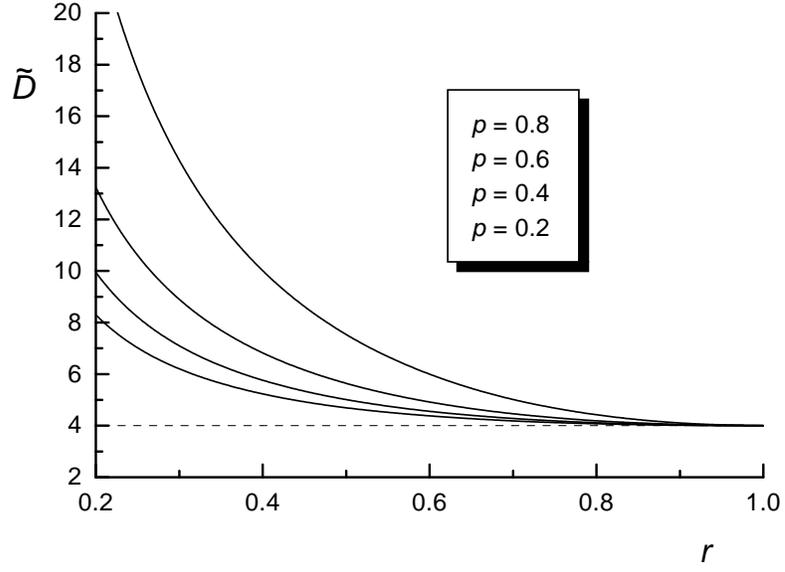,bbllx=14pt,bblly=440pt,bburx=540pt,bbury=850pt,width=0.6\linewidth}}
\vspace{0.9cm}
\caption{$\tilde D$ as a function of the multiplicity ratio
$r\equiv \langle N_- \rangle / \langle N_+ \rangle$  for several
values of $p$. The top line corresponds to $p=0.8$,  
the next one to $p=0.6$, etc.}
\end{figure}

\vspace{-0.7cm}
\begin{figure}
\centerline{\epsfig{file=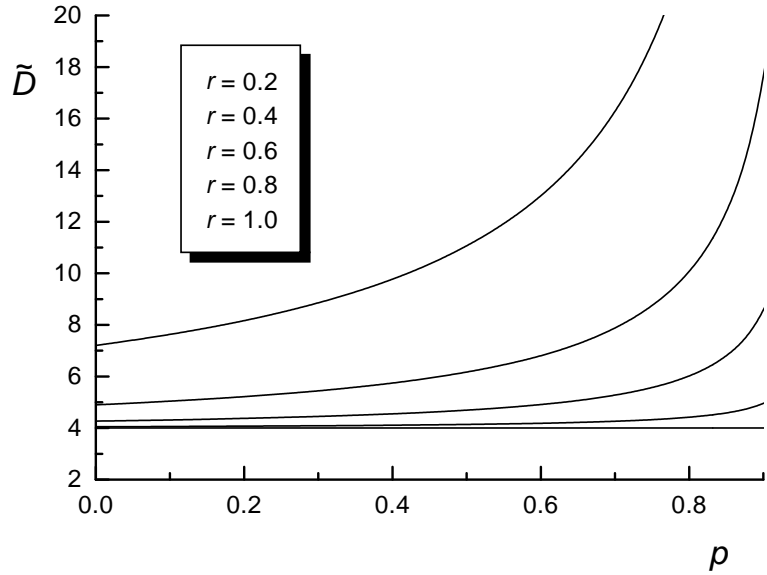,bbllx=14pt,bblly=440pt,bburx=540pt,bbury=850
pt,width=0.6\linewidth}}
\vspace{0.9cm}
\caption{$\tilde D$ as a function of the particle registration
probability $p$ for several values of
$r\equiv \langle N_- \rangle / \langle N_+ \rangle$.
The top line corresponds to $r=0.2$, the next one
to $r=0.4$, etc. }
\end{figure}

\newpage

\section{Summary and conclusions}
 
We have shown that the $\Phi-$measure, which has been successfully
applied to the transverse momentum fluctuations, can be also used
to study charge fluctuations. $\Phi$ is insensitive to the
collision centrality and sensitive to the dynamics. If one is
interested not only in the dynamical fluctuations but in the absolute
value the fluctuations the $\Gamma-$measure, which is related to
$\Phi$, can be applied. We have established the relationships
between $\Phi$, $\Gamma$ and several charge fluctuation measures
proposed by other authors. The measure $D$ \cite{Jeo00,Ble00} has
been shown to be equivalent to $\Gamma$ but for small fluctuations only.

The charge fluctuations have been analyzed within a `background'
model where the particles are produced statistically being
constrained by charge conservation. The effects of the finite acceptance
of experimental apparatus and of the detection inefficiency are
schematically taken into account. The model's results depend solely
on the two parameters: $p$ which is a fraction of all produced
particles taken to the analysis and $r$ being the ratio
$\langle N_- \rangle / \langle N_+ \rangle$. The `background' model's
predictions are close to the PHENIX \cite{Adc02} and preliminary STAR
\cite{Vol01} results on electrical charge fluctuations. Thus,
there is little space for the dynamical effects such as freezing of
the fluctuations generated at the quark-gluon plasma phase
\cite{Jeo00,Asa00}.

At the end we have considered the measures which are supposed to be
free of the effects of charge conservation and incomplete acceptance. 
We have shown within our `background' model that when $r$ is
noticeably smaller than unity, as in the case of the baryonic
charge fluctuations, the simple corrections proposed by other
authors do not work properly. 

The data on the charge fluctuations \cite{Rol98,Adc02,Vol01}
show that the dynamical phenomena do not contribute significantly 
to the observed fluctuations. Thus, a proper choice of a statistical
tool for data analysis is important to quantify and interpret the small
effects of interest. It is also important to carefully eliminate
trivial effects as those caused by the charge conservation and  
non-vanishing net charge. The results presented here contribute
to these goals.

\begin{acknowledgements}

I am very grateful to Sergei Voloshin for the correspondence that 
initiated this work. I am indebted to him, Marek Ga\' zdzicki 
and Jacek Zaranek for critical comments on the manuscript. My
thanks also go to Joakim Nystrand and Evert Stenlund for a
discussion on the PHENIX data.

\end{acknowledgements}

\end{document}